# A Multi-Tenant System for 5/6G Testbed as-a-Service

Raffaele Bolla
*DITEN - University of Genoa*
Genoa, Italy
raffaele.bolla@unige.it

Roberto Bruschi
*DITEN - University of Genoa*
Genoa, Italy
roberto.bruschi@unige.it

Chiara Lombardo
*CNIT – S2N National Lab*
Genoa, Italy
chiara.lombardo@cnit.it

Sergio Mangialardi
*CNIT – S2N National Lab*
Genoa, Italy
sergio.mangialardi@cnit.it

Alireza Mohammadpour
*CNIT – S2N National Lab*
Genoa, Italy
alireza.mohammadpour@cnit.it

Ramin Rabbani
*DITEN - University of Genoa*
Genoa, Italy
ramin.rabbani@edu.unige.it

Beatrice Siccardi
*DITEN - University of Genoa*
Genoa, Italy
beatrice.siccardi@edu.unige.it

*Abstract*—In order to fulfill the stringent requirements and fast advancements of 5G and beyond applications, it is inevitable to develop research/industrial testbeds to examine the different proposed innovative features of 5G and beyond. In this paper, we propose a testbed including 5G and beyond technologies by combining open-source solutions and our developed Network Services/Functions to enhance the ETSI-NFV MANO framework. Our testbed contains an automation framework to reduce both run time and setup complexity. It provides various services: Metal as a Service (MaaS), Infrastructure as a Service (IaaS), Platform as a Service (PaaS), different Network Functions (NFs), and Network Services (NSs), under the context of full automation of End-to-End (E2E) NSs on top of the services provided by ETSI.

## I. INTRODUCTION

The upcoming Fifth-Generation (5G) brought with it new innovative services and models that play a key role in the digital transformation towards a new hyper-connected society [1], [2]. 5G-ready applications can be composed of independent, cloud-native *"microservices"* [3] running on individual execution environments (e.g., Virtual Machines (VMs), or containers) deployed across multiple facilities and connected via network slices, which represent logical End-to-End (E2E) networks, composed of Virtual Network Functions (VNFs), providing specific 5G Network Services (NSs). Physical infrastructure resources are limited and valuable, especially when data traffic demands from the different operators increase. Therefore, network slicing, a development of network sharing, can be considered as a conventional solution [4] in which multiple operators can share infrastructure resources according to their plans for allocating resources. Network Function Virtualization (NFV), Software Defined Networking (SDN), and Cloud computing are considered the three enabling technologies for implementing network slicing in 5G [5]. In this regard, research communities in academia and industry, focusing on examining different features of 5G and beyond, are developed. Their work leads to proposing some prototype system implementations of individual parts of the mobile network architecture in 5G, which is known as a research testbed. Research testbeds provide the opportunity to evaluate and improve network performances. Additionally, they keep the cost of network deployment low, and their functionalities are almost identical to those of actual networks. They can also leverage the availability of open-source software packages for investigating and creating novel solutions applied in 5G and beyond [6]. In this paper, we combined the open-source solutions and our developed NSs/Network Functions (NFs) to enhance the ETSI-NFV MANO framework [7]. By leveraging on an automation framework to reduce both run time and setup complexity, our testbed can provide Metal as a Service (MaaS), Infrastructure as a Service (IaaS), Platform as a Service (PaaS), Testbed as a Service (TaaS), different NFs, and NSs under the context of Zero-touch network and Service Management (ZSM) framework targeting the full automation of E2E NSs on top of the services provided by ETSI [1].

This paper is organized as follows. Section II proposes a state of the art about different proposed testbeds for 5G and beyond and analyzes their pros and cons. Section III introduces our testbed, the CNIT-UniGE 5/6G Testbed, with a particular focus on its main software component: the Operation Support System (OSS). The services/functions our testbed offers together with the projects that it hosted/is going to host are mentioned as well. Finally, conclusions of this study are drawn in Section IV.

## II. STATE OF THE ART

Testbeds are used to examine different features of the new technologies, such as 5/6G, and to evaluate network performance under various scenarios. Testbeds are used in academic and industrial research widely. Most of the practical research works in 5G and beyond are carried out in research testbeds, such as developing prototype system implementation of individual parts of the mobile network architectures. Applying network slicing capabilities in the testbed is a challenging task requiring specific resources such as network equipment and the ability to configure and program the physical infras-tructure. The physical and virtual components of a network slicing testbed should provide appropriate performance for the requested services. There are plenty of 5/6G testbeds proposed for applying some specific services using network slicing. In [8] the proposed testbed is implemented in several containers to virtualize both Evolved Packet Core (EPC) and eNodeBs. It is created on a cloud infrastructure based on OpenStack and the proposed services can be created via network slices. The testbed in [9] follows the Smart E2E Massive IoT Interoperability, Connectivity, and Security (SEMIoTICS) architecture to create a 5G platform based on SDN, NFV, and Multi-access Edge Computing (MEC). In POSENS platform [10] different NSs in the network layer are chained by MANO to create network slices. It also applies efficient resource utilization for creating independent and customizable E2E slices. UPC University Testbed [11] platform implements Radio Access Network (RAN) slicing via RESTful APIs automatically. The testbed applies the slice-aware policy in Radio Resource Management (RRM) functionalities for admission control and scheduling processes. There are also plenty of testbeds that are available and in [5] a review of the available testbeds and their architecture is proposed.

In this paper, we propose the testbed jointly developed by National Inter-university Consortium for Telecommunications (CNIT) and University of Genoa (UniGE). Our testbed's main goals are to provide, through a blend of both hardware and software components, full bare-metal control of the hardware and customizable 5/6G networks.

## III. THE CNIT-UNIGE 5/6G TESTBED

The CNIT-UniGE 5/6G testbed, is part of the Scientific Large-Scale Infrastructure for Computing/Communication Experimental Studies (SLICES) [12] project, which has been selected to be part of the 2021 roadmap of the European Strategy Forum on Research Infrastructures (ESFRI) [13]. Our testbed is a multi-layered hardware and software facility containing both general- and special-purpose equipment for the full bare-metal control of the hardware to build customizable 5/6G environments, to completely monitor the system, and to expose remote access to resources. It is specifically conceived to host multiple isolated tenant spaces (e.g., as project environments) that can emulate complete 5G network environments, as well as to manage and configure their respective physical/virtual resources through a MaaS or IaaS approach.

The hardware resources of the testbed are listed as follows. *1)* Computing resources include 22 high-end servers (700 cores, 2.5TB RAM, local and central SSDs/SAS storage >100TB). The servers support remote management systems (e.g., IPMI, Intel AMT, HP iLO) to control their power states and access internal sensors and probes even in the down state or standby state. *2)* Networking resources composed of 8 high-speed switches (918 ports from 1GbE to 100GbE) and 1 P4 Tofino-based switch. *3)* 2 AmariCallbox 5G devices gNodeB MIMO 4x4 [14] and 3 LTE+ eNodeBs (USRP-based) as base stations. *4)* User Equipment (UE) including 8 Commercial 5G smartphones, 4 5G Rel. 15 modem routers, and 1 Amari 5G Simbox-64 [15]. *5)* Power consumption probes containing 4 Raritan smart power distribution to meter, expose energy consumption, and schedule the power cycle of each outlet (device). Moreover, there is an Agilent 2300 modular system equipped with 2 multi-meters and 2 high-speed Digital Analog acQuisition (DAQ) systems to measure internal server power consumption. The testbed is also equipped with smart UPS systems allowing to cap energy and power provided to devices. They can be exploited to emulate energy-constrained environments. *6)* High-precision time synchronization: all the servers, computing units, gNodeBs, and measurement devices are synchronized with an IEEE 1588v2 GPS Grandmaster clock with an accuracy of approximately 20 ns.

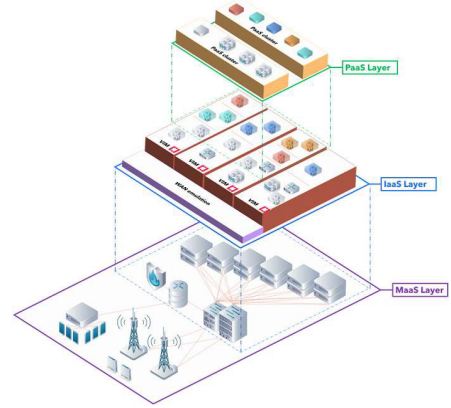

Fig. 1. The three levels of programmability of the CNIT-UniGE 5/6G testbed: MaaS (purple), IaaS (blue) and PaaS (green).

Regarding the software components, we developed a custom Operation Support System that provides, through RESTful APIs, on one hand, the management and terraforming of bare-metal resources (i.e., physical servers, hardware networking equipment and physical devices (i.e., g/eNodeBs)) to create environments with different programmability levels (e.g, MaaS, IaaS or PaaS); on the other hand, the management of the lifecycle of NFV services to provide suitable connectivity to vertical application components and UEs in a fully automated and zero-touch fashion.

The three programmability levels provided by our testbed are shown in Fig. 1. The first layer of programmability is MaaS which offers bare-metal resources. The two top layers are achieved by virtualizing the aforementioned resources. In detail, the testbed offers both IaaS and PaaS programmability: the former offers the possibility to deploy VMs by using OpenStack which acts as the Virtual Infrastructure Manager (VIM) in the ETSI NFV Architectural Framework [7]; the latter offers the deployment of containerized applications on a Kubernetes cluster running on VMs. These top layers are necessary to support the cloud-native nature of the 5G Core Network (5GC) [6], whose NFs should be deployed as VNFs (based on VMs) or Container Network Functions (CNFs). Finally, it is worth mentioning that the testbed physical/virtual resources can be accessed remotely through the use of a Virtual Private Network (VPN); this can be also exploited to realize site-to-site connections with other testbeds.

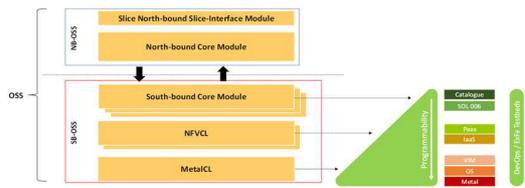

Fig. 2. The OSS architecture.

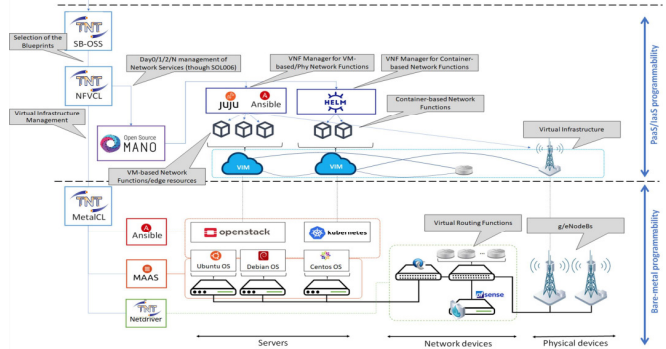

Fig. 3. The SB-OSS architecture.

A. The proposed Operation Support System (OSS)

The proposed Operation Support System is the heart of the software stack of the testbed. It was developed within the MATILDA [16] project first, and then the 5G-INDUCE one [17]. In particular, 5G-INDUCE aims to exploit, self-optimize, automate, and make as simple as possible the management of Network Applications (NetApps) onto 5G and beyond (B5G) infrastructures. Our OSS is in charge of managing all functions and operations required for the NetApp placement over edge computing facilities and for its connection to a (properly configured) network slice, as well as maintaining the information on all the data regarding the deployed applications, network services, and available infrastructure resources.

A crucial point of our current OSS is that it supports any NFV levels: NSs can be composed by mixing VNFs realized not only with IaaS resources but also with cloud-native containers over PaaS platforms (e.g., 3GPP 5GC NFs over Kubernetes clusters), and physical programmable/configurable devices (e.g., gNodeBs, eNodeBs, etc.). Moreover, the OSS is designed according to a highly modular architecture: all the software services are state-of-the-art cloud-native software, i.e., stateless services (or more precisely services with a state maintained in an external database, namely MongoDB), inherently parallelizable, and ready to be integrated with the service-mesh paradigm.

As shown in Fig. 2, the OSS is split into two modules: the North-Bound OSS (NB-OSS) and the South-Bound OSS (SB-OSS). The NB-OSS represents the natural extension of the Network Application Orchestrator (NAO) in the Telecom Service Provider domain and is meant to front-face the NAO by managing slice negotiations for NetApps, and to maintain metadata (e.g., coverage area served, operational capabilities, etc.) of one or multiple onboarded SB-OSS modules. The NB-OSS itself is composed of two main services: the Slice-Interface and the North-Bound (NB) Core modules. The Slice-Interface service is responsible for the interaction between the NAO and the OSS. In particular, it oversees the messages to be exchanged in order to request the creation of a slice, including the computational, networking and Quality of Service (QoS) requirements, as well as for the slice lifecycle management up to its termination. The NB Core module is in charge of two main tasks: onboarding SB-OSS instances, and suitably process and propagate slicing requests/replies between the Slice-Interface service (and then the NAO) and the relevant SB-OSS(es).

The SB-OSS is meant to be an adaptation software stack to onboard heterogeneous 5G infrastructures (and, therefore, it may fall at least partially into the domain of Telecom Infrastructure Providers); moreover, it is deployed once per each different administrative networking/computing resource domain onboarded onto the OSS. To reflect the different programmability levels exposed by such administrative domains, the SB-OSS has been designed as a chain of software services that can be selectively activated to gain access to various programmability levels. As depicted in Fig. 3, the SB-OSS is composed of three services: the South-Bound (SB) Core module, the NFV Convergence Layer (NFVCL) and the Metal Convergence Layer (MetalCL). The SB Core module is devoted to process the slice instantiation, modification, de-instantiation requests, and related resources. This service is the key component for providing adaptive programmability: if the NFVCL and the MetalCL services are available, the SB Core module can request them the setup or change of new or existing network slices/services, and of infrastructure resources (e.g. of OpenStack VIM instances and of the physical servers composing it). The NFVCL manages the lifecycle of NFV services to provide suitable connectivity to NetApp components and UEs in a fully automated and zero-touch fashion. Finally, the MetalCL is dedicated to manage and terraform bare-metal resources (i.e., physical servers and hardware network equipment) available in the testbed to create IaaS/PaaS environments compliant with the 5G-platform needs.

*1) The NFV Convergence Layer (NFVCL):* The NFV Convergence Layer [1] fully drives NFV service orchestration along all the lifecycle phases. The objective of NFVCL is providing a level of abstraction for the flexible and high-level management of the complete lifecycle orchestration of NSs, VNFs, Kubernetes-based Network Functions (KNFs), and Physical Network Functions (PNFs) instantiated in the 5G infrastructure [18]. PNFs are needed in order to support real base stations. More precisely, the NFVCL is able to build and to dynamically manage complete network environments

(e.g., a 5G core network with different Next Generation (NG)-RANs) by composing and orchestrating through an NFV Orchestrator (NFVO) (usually Open Source MANO (OSM) [19]) multiple NFV services in a joint and coherent fashion. This logical group of NFV services realizing a certain network environment is managed within the NFVCL by a *"blueprint"* which is a software plugin allowing the dynamic composition and the coordinated Day-0/1/2/N management of several NSs (each composed of VNFs, KNFs and PNFs) realizing together a complex network functionality (e.g., a 5G system). The NFVCL was designed to support multiple types of *blueprints*, and multiple instances per type. The number and the type of NFV services (e.g., the number and the specific implementation of NG-RAN NFV services in a 5G network) to be used is dynamically selected by the *blueprint* according to the request parameters (e.g., the geographical coverage requested for the slice). In a nutshell, the NFVCL is in charge, upon request from the SB Core module, to select the most suitable *blueprints* to be instantiated or the *blueprint* instances to be updated/reconfigured to satisfy the performance and functional requirements included in the slice request, and then to provide feedback to the SB Core module when all the NFV services in the *blueprint* are ready and all the relevant VNFs, KNFs, and PNFs have been successfully configured.

Moreover, the NFVCL is in charge of producing configuration commands and files to be passed to the NFs through Day-2 operations towards their VNF Managers (VNFMs), and it is able to retrieve operation output, logs, and performance metrics from them. Regarding the VNFMs, it can be noticed from Fig. 3 that three are used. As far as VNFs and PNFs are concerned, Juju and Ansible are used. Initially, the VNFM responsible for the service instantiation was realized by using Juju; this choice meant that a newly developed charm was required for each onboarded NF, with a huge impact on the instantiation/onboarding time [20]. For this reason, a new type of VNFM was realized, called FlexCharm, which allows to minimize onboarding time thanks to its generalized structure: with Flex, the charm is the same for each NF, and it contains an Ansible instance that, upon reception of instructions in YAML format, identifies the lists of playbooks and files to be applied. This allows also the NFVCL to build configuration with a uniform approach, since for each NF, configuration files and commands are produced with the same FlexCharm interface. CNFs, on the contrary, do not use the aforementioned FlexCharm VNFM; they rely on the Helm tool which is integrated into the NFVO (i.e., OSM).

The NFVCL layer was initially introduced in the architecture for two reasons: on one hand, it was designed to support different NFVOs without the need to intervene on the OSS structure (since it implements standard ETSI NFV-SOL 004 [21], 005 [22] and 006 [23] APIs); on the other hand, it allowed to cope with some shortcomings of ETSI NFV specifications, as explained herein. The Network Service Descriptors (NSDs) implemented in ETSI NFV are static, as the number of VNFs and the related internal/external connections is pre-determined in accordance with the VNF Descriptors (VNFDs). This makes

```
{
    "type": "K8s",
    "config": {
        "version": "1.24",
        "network_endpoints": {
            "mgt": "control",
            "data_nets": [
                {
                    "mode": "layer2",
                    "net_name": "control"
                }
            ]
        }
    },
    "areas": [
        {
            "id": 3,
            "core": true,
            "workers_replica": 1
        }
    ]
}
```

Fig. 4. The body of a POST message to create a Kubernetes cluster.

it impossible, for example, to perform network service internal topology change on any of the functions without destroying and re-deploying the whole service, let alone to modify the in-life operations on-the-fly upon changes in the related slice. The NFVCL has been enhanced by optimizing the procedure for the creation of the VNF/NS packages to be onboarded on the NFVO. This new capability is provided by the full implementation of ETSI SOL 006 specification.

The NFVCL service has also evolved to support some basic VIM terraforming operations (e.g. creation of a network) and to maintain a topology of the virtual network infrastructure (which can be onboarded as static or dynamically created/modified). Networks in the topology can be used as endpoints for NFV services.

Finally, the NFVCL is an intent-based server following the OpenAPI standard. It uses the FastAPI Python library that natively supports the aforementioned standard. Therefore, the NFVCL follows the CRUD paradigm in which each action (create, read, update, delete) is mapped onto an HTTP message (POST, GET, PUT, DELETE). In order to interact with the NFVCL, one can either use an *ad-hoc* web-based Graphical User Interface (GUI) or directly send HTTP messages. An example of the body of such an HTTP message is provided in Fig. 4. This message requests the NFVCL to create a Kubernetes cluster with only one worker, located in the area whose id is equal to 3; we provide a geographical separation into numbered areas in order to support edge computing through simulating the presence of edge datacenters in different geographic parts. Moreover, it is worth noting that two network endpoints should be specified: one for management and one for data forwarding purposes.

*2) The Metal Convergence Layer (MetalCL):* The MetalCL is devoted to the management and terraforming of VIMs, Operating Systems (OSs) and bare-metal resources available in a testbed. In the presence of this service, physical servers and hardware networking equipment, as well as their OSs, can be dynamically managed on demand, in a similar way (though, on different time scales) as NFV services. As depicted in Fig. 3, the MetalCL connects with three external blocks that can be selectively enabled depending on the programmability level of the administrative infrastructure domain: the NetDriver, the MaaS server, and the Ansible engine.

The NetDriver is used to create proper overlay networks and to manage the interconnectivity among servers as needed (i.e., to create proper networks for data- and control-plane for hosting OpenStack, Kubernetes, etc.). It uses the LLDP protocol to perform an automatic discovery of the physical topology, and it is meant to access the command line interface or to use the RESTful interface of physical networking devices in the computing facility (e.g., interconnection layer-2 managed switches, with VLAN or OpenFlow support; routers, optionally with the support of virtual routing functions; and firewalls).

TABLE I
SERVICES/FUNCTIONS OFFERED BY OUR TESTBED.

| Services/Functions | Description |
|---|---|
| EPC, 5Gc, IMS, eNodeB, gNodeB | provided by the Amari Callbox. |
| Free5GC | an open-source implementation of the 5GC [24]. Our testbed supports both a VM-based and a container-based implementation of such a network. |
| Open5Gs (EPC + 5GC) | an open-source implementation of 5GC and EPC [25]. Our testbed supports both a VM-based and a container-based implementation of such networks. |
| NextEPC | an open-source implementation of the 4G/5G 3GPP core network [26]. |
| OpenAirInterface | an open-source implementation of the 3GPP stack: the RAN (eNB, gNB and 4G, 5G UE) as well as the core network (EPC and 5GC) [27]. |
| VyOS router | an open-source router platform [28]. |
| Kubernetes | an open-source platform for managing containerized workloads and services [29]. Our testbed supports the installation of Kubernetes clusters in two ways: on VMs and bare-metal. |
| ELK | an open-source stack used as a log analytics tool [30]. |
| OpenWrt | an open-source project for embedded operating systems based on Linux, primarily used on embedded devices to route network traffic [31]. |
| OpenVSwitch | an open-source implementation of a distributed multilayer software switch. The main purpose of OpenVSwitch is to provide a switching stack for hardware virtualization environments. |
| S1 Bypass | a custom VNF related to 4G and based on Data Plane Development Kit (DPDK). It filters packets coming from eNodeBs and, according to customized rules, bypasses the S-Gateway in order to steer traffic to the desired edge data center. For further detail please refer to [1]. |
| UE emulator | provided by the Amari 5G Simbox. |
| T-Rex Traffic generator | a software traffic generator as a Service to measure the performance of the virtual networks used in a fully virtual environment based on the open-source T-REX project by Cisco; it was presented in [32]. |
| Virtual Object OMA Lwm2m | an implementation of a device management protocol designed for sensor networks and the demands of a Machine-to-Machine (M2M) environment. |
| UERANSIM | an open-source 5G UE and RAN (gNodeB) simulator [33]. |
| Desktop as a Service (also for UEs) | it provides secure DaaS through HTTPS protocol using Guacamole [34]. |
| Prometheus | an open-source monitoring and alerting solution for collecting and aggregating metrics as time series [35]. |
| Packet delay simulator | a custom network delay simulator based on DPDK that emulates typical latency delays of Wide Area Network (WAN) links. |

The Metal as a Service server is an open-source project developed by Canonical [36]. It allows the discovery, commission, deployment, and dynamic reconfiguration of a large network of individual bare-metal servers. MaaS is able to catalogue bare-metal servers in a data center, to control their power states, and to install (on demand and as a Service) almost any operating system. MaaS exposes a complete set of RESTful APIs, which are consumed by the MetalCL to trigger any changes at the bare-metal level.

At last, MetalCL makes use of the Ansible engine to drive any software installation, application and OS reconfigurations over the bare-metal servers installed by MaaS. This engine is the one that provides the MetalCL with the capability of installing software dependencies and to install and to manage the lifecycle, in a zero-touch fashion, of complex and distributed software (e.g., OpenStack or Kubernetes) over one or more servers.

Finally, the MetalCL is, as well as the NFVCL, a REST-based server that can be accessed by a web-based GUI that is shown in Fig. 5.

### B. Proposed Network Services/Functions

As previously mentioned, the main goal of the proposed testbed is to provide custom 5/6G environments to verticals; however, in order to support several use cases, other services/functions should be supported as well. Such offered functions in particular, can have three forms: VNFs, KNFs and PNFs. Table I presents a list of the Network Services/Functions that the testbed provides. It is worth noting that most of the offered services/functions are based on open-source tools.

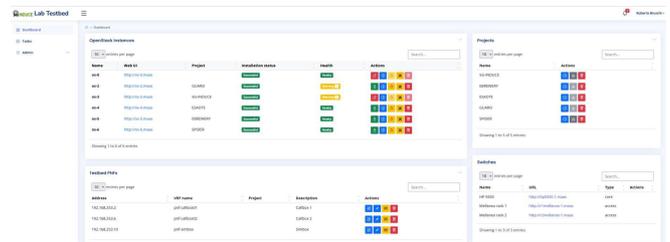

Fig. 5. The MetalCL GUI

### C. Hosted European Projects

The proposed testbed has been, and currently is, part of several European projects. In detail, it supports several projects at the same time since the appropriate required services (e.g., MaaS, IaaS, PaaS, NSs, etc.) are fully isolated. In the following, a non-exhaustive list of past, current, and future projects in which the testbed was/is involved is provided. All the following projects were/are funded by the European Commission. Among the past projects, it is worth mentioning MATILDA [16], whose goal was to design and implement a holistic 5G E2E services operational framework tackling the lifecycle, design, development, and orchestration of 5G-ready applications and 5G network services over programmable infrastructure; and SPIDER [37], whose aim was to develop a replicable cyber range platform for the telecommunications domain and its fifth generation, offering cybersecurity emulation, training, and investment decision support.

Currently, the most significant projects are: 5G-INDUCE [17], mentioned previously in the paper; NEPHELE, whose goal is to enable the efficient, reliable, and secure E2E orchestration of hyper-distributed applications over programmable infrastructure that is spanning across the compute continuum from Cloud-to-Edge-to-IoT; and SLICES [12], whose aim is to provide researchers with a common flexible infrastructure to support research reproducibility and experiment repeatability. Finally, among future projects, it is worth mentioning 6Green and HORSE. The former's aim is to conceive, design, and realize an innovative service-based and holistic ecosystem, able to extend the communication infrastructure into a sustainable, interconnected, greener E2E intercompute system. The latter proposes a novel human-centric, open-source, green, sustainable, coordinated provisioning and protection evolutionary platform, which can inclusively yet seamlessly combine advancements in several domains, as they get added to the system (e.g., predictive threats detection, AI-based techniques, etc.).

## IV. Conclusions

This paper proposed a research/industrial testbed able to prove a wide variety of services proposed in 5G and beyond such as MaaS, IaaS, PaaS, DaaS, various NSs, and NFs through an automation process by utilizing the proposed customized OSS structure. It provides the ability to create, modify, and delete NSs through RESTful APIs. The testbed has already been hosting multiple European research projects since it is compatible with new upcoming technologies. The performance and functionalities of the testbed are developing using the updated features and innovative services.

## V. Acknowledgement

This work has been supported by the Horizon Research and Innovation Action NEPHELE (Grant Agreement no. 101070487), SLICES-SC (Grant Agreement no. 101008468), HORSE (Grant Agreement no. 101096342), 6Green (Grant Agreement no. 101096925), and by the Horizon Innovation Action 5G-INDUCE (Grant Agreement no. 101016941).